\documentclass{article}[10pt,english]
\usepackage{graphicx}
\usepackage{pstcol,times,color,graphicx,graphics,pstricks}
\def\[{\begin{equation}}
\def\]{\end{equation}}
\def\ex{\mbox{e}}
\begin{document}
\title{Analytical regularisation for confined quantum fields between parallel surfaces}
\author{  F C Santos \footnote{e-mail:
filadelf@if.ufrj.br}, A C Tort
\footnote{e-mail: tort@if.ufrj.br.}\\
Departamento de F\'{\i}sica Te\'{o}rica - Instituto de F\'{\i}sica
\\
Universidade Federal do Rio de Janeiro\\
Caixa Postal 68.528; CEP 21941-972 Rio de Janeiro, Brazil\\
and \\
E Elizalde\footnote{e-mail: elizalde@ieec.uab.es} \\
Instituto de Ciencias del Espacio (CSIC) \\ \&
Institut d'Estudis Espacials de Catalunya (IEEC/CSIC) \\
Campus UAB, Facultat de Ci\`encies, Torre C5-Parell-2a planta \\
E-08193 Bellaterra (Barcelona) Spain} \maketitle
\begin{abstract}
We review a simple technique for evaluating the vacuum energy stemming
from non-trivial boundary conditions and present results for the
Casimir energy of a massive fermionic field confined in a
$d+1$-dimensional slab-bag and the effect of a uniform magnetic
field on the vacuum energy of confined massive bosonic and fermionic
fields. New results concerning the Casimir energy and the evaluation
of the rate of creation of quanta in $\kappa$-deformed theories are
presented.
\end{abstract}
\section{Introduction}
In quantum field theory under external conditions, the macroscopically observable zero-point energy shift is defined as the regularised difference between the vacuum
expectation value of the hamiltonian with and without
the presence of the external conditions. Under certain assumptions, external conditions can be approximated by boundary conditions and one-loop calculations of the energy shift lead to the Casimir effect \cite{Casimir48} -- see Ref. \cite{BMohideenM2001} for the most recent review on this subject. In the evaluation of the zero point energies some configurations which depend on
the nature of the particular quantum field (scalar, spinorial, etc.), the type of spacetime manifold and its dimensionality, and the specific boundary condition imposed on the quantum field on certain surfaces, may lead to rather complex spectra. The heart of the matter in these calculations is the evaluation of the spectral sum that results at the one-loop level from the shift of the zero-point energy. This evaluation requires regularisation and renormalisation, and recipes for accomplishing this task range from the relatively simple cutoff method, employed by Casimir himself \cite{Casimir48}, to a number of powerful and
generalised zeta function techniques \cite{Elisalde94}. Contour
integral representations of spectral sums are a great improvement
in the techniques of evaluating zero-point energies and have been employed before \cite{relatedpapers}. A simple technique for evaluating vacuum energy stemming from  non-trivial boundary conditions based on Cauchy's integral formula and the Mittag-Leffler expansion theorem was proposed in \cite{ ST2002, IJMPA1003}. The method was employed in a number of cases, in particular it was applied to calculate the Casimir energy of a massive fermionic field confined in a $d+1$-dimensional slab-bag \cite{IJMPA1003} and to calculate the effect of a uniform magnetic field on the vacuum energy of confined massive bosonic and fermionic fields \cite{JPA2002}. Recently this technique was applied to evaluate the Casimir energy of $\kappa$-deformed theories under Dirichlet boundary onditions \cite{POSWC2004}. Here we review these cases and extend the method to the evaluation of the rate of creation of quanta in $\kappa$-deformed theories.
\section{A general formula for the regularised vacuum energy shift}
For a field theory in $d+1$ flat spacetime under boundary conditions imposed on  two parallel planes of area equal to $L^{d-1}$ which are kept at a fixed distance $\ell$ from each other, and with $L\gg \ell$, the non-regularised vacuum energy is given by
\[\label{E0}
E_0 \left( d \right) = \alpha \left( d \right)\frac{{L^{d - 1} }}{2}\int {\sum\limits_{n} {\frac{{d^{d - 1} p_ \bot  }}{{\left( {2\pi } \right)^{d - 1} }}} } \Omega _n
\]
where $\alpha (d)$ is a dimensionless factor that takes into account the internal degrees of freedom and
\[
\Omega _n  = \sqrt {p_ \bot ^2  + \frac{{\lambda _n ^2 }}{\ell } + m^2 }
\]
with $ p_ \bot ^2  = p_1^2  + p_2^2  + p_3^2  +  \cdots  + p_{d - 1}^2 $
and $m$ is the mass of an elementary excitation of the field and the $\lambda_n$ are the real roots of the equation determined by the boundary conditions. Consider now
\[\label{eq5}
\sum\limits_n {\Omega _n  =  - \oint\limits_\Gamma  {\frac{{dq}}{{2\pi }}\sum\limits_n {\frac{{2q^2 }}{{q^2  + \Omega _n^2 }}} } }
\]
which follows from Cauchy's integral formula. The curve $\Gamma$ is a Jordan curve on the complex $q$-plane with $\mbox{Im} q > 0$ that we choose to be a semicircle with very large radius the diameter of which is the entire real axis of the auxiliary variable $q$.
%
%\begin{figure}[!h]
%\begin{center}
%\begin{pspicture}(-4,-1)(4,5)
%\showgrid
%\psset{arrowsize=0.2 2}
%\psline[linewidth=0.5pt]{->}(-4,0)(4,0)
%\rput(3,-0.5){${\mbox{Re q}}$}
%\psline[linewidth=0.5pt]{->}(0,-1)(0,4)
%\rput(-0.65,3.5){${\mbox{Im q}}$}
%\psarc[linewidth=0.15mm]{-}(0,0){3}{0}{180}
%\psdots(0,0.35)
%\psdots(0,1.75)
%\psdots(0,2.5)
%\end{pspicture}
%\caption{My caption.}
%\label{My label}
%\end{center}
%\end{figure}
%
Defining
\[
z = \ell \sqrt {q^2  + p_ \bot ^2  + m^2 }
\]
we can write
\[
\sum\limits_n {\Omega _n  =  - } \oint\limits_\Gamma  {\frac{{dq}}{{2\pi }}} \frac{{\ell ^2 q^2 }}{z}\sum\limits_n {\frac{{2z}}{{z^2  + \lambda _n^2 }}}
\]
This summation can be performed in the following way: Let $G(z)$ be
a complex function of the complex variable $z$ symmetrical on the
real axis and suppose its roots $\lambda_n$ are simple, nonzero and
symmetrical with respect to the origen. Notice that if $z=0$ is a
root of $G$ we can divide $G$ by some suitable power eliminating
this root from the set of roots without introducing new
singularities. Since the roots are symmetrical we can order them in
such a way that $\lambda_n=-\lambda_{-n}$ with $n=\pm 1, \pm 2,...$.
Defining the auxiliary function $J(z)$ by
\[
J\left( z \right) = \sum^{\;\;\;\;\;\;\;\;\prime}\limits_{n} {\frac{1}{{z - i\lambda _n }}}
\]
where the prime indicates the the term $n=0$ is to be omitted. This function is meromorphic with simple poles at $i\lambda_n$ and residua equal to one. Defining the function $K(z)=G(iz)$ we see that the function $K^{\,\prime}(z)/K(z)$ has simple poles at $z=i\lambda_n$ with residua also equal to one. Due to the symmetry of the roots  we can write
\[
J\left( z \right) = \frac{{K'\left( z \right)}}{{K\left( z \right)}} \Rightarrow \frac{{K'\left( z \right)}}{{K\left( z \right)}} = \sum\limits_{} {\frac{{2z}}{{z^2  + \lambda _n^2 }}}
\]
We can make use of this identity and of equation (\ref{eq5}) to recast equation (\ref{E0}) into the form
\[
E_0 \left( d \right) =  - \alpha \left( d \right)\frac{{L^{d - 1} }}{2}\int {\frac{{d^{d - 1} p_ \bot  }}{{\left( {2\pi } \right)^{d - 1} }}} \oint\limits_\Gamma  {\frac{{dq}}{{2\pi }}} q\frac{{d\ln \left[ {K\left( z \right)} \right]}}{{dq}}
\]
Integarting by parts we obtain
\[\label{shift}
E_0 \left( d \right) = \alpha \left( d \right)\frac{{L^{d - 1} }}{2}\int {\frac{{d^{d - 1} p_ \bot  }}{{\left( {2\pi } \right)^{d - 1} }}} \int {\frac{{dq}}{{2\pi }}\ln \left[ {K\left( z \right)} \right]}
\]
At this point it is possible to simplify this last expression by decomposing the function $ K(z)$ into two parts, namely $
K\left( z \right) = K_1 \left( z \right) + K_2 \left( z \right)$ where $K_1(z)$ contains all terms whose integrals diverge when $\mbox{Re} > 0$ and $K_2(z)$ contains all terms whose integrals diverge when $\mbox{Re} < 0$; also $K_1(z)=K_2(-z)$. In this way the integral in equation (\ref{shift}) can be written as
%
%$$
%I=\int\frac{dq}{2\pi}\ln\left[K(z)\right]
%$$
%
\begin{eqnarray}
I &=& \int_{\Gamma_1}\frac{dq}{2\pi}\ln\left[K_1(z)\right] + \int_{\Gamma_1}\frac{dq}{2\pi}\ln\left[1+\frac{K_2(z)}{K_1(z)}\right]\nonumber \\
&+& \int_{\Gamma_2}\frac{dq}{2\pi}\ln\left[K_1(z)\right]+\int_{\Gamma_2}\frac{dq}{2\pi}\ln\left[1+\frac{K_1(z)}{K_2(z)}\right]
\end{eqnarray}
The regularised vacuum energy shift is then
\[
E_0 \left( z \right) = \alpha \left( d \right)\frac{{L^{d - 1} }}{2}\int {\frac{{d^d p}}{{\left( {2\pi } \right)^d }}\ln \left[ {1 + \frac{{K_1 \left( z \right)}}{{K_2 \left( z \right)}}} \right]}
\]
%Making use of the identity
%
%\[
%\int {d^d p\,F\left( p \right) = \frac{{2\pi ^{d/2} }}{{\Gamma \left( {d/2} \right)}}\int\limits_0^\infty  {dp\,p^{d - 1} F\left( p \right)} }
%\]
Rewriting  $z=\sqrt{p^2+m^2}$ with $p:=x/\ell$ and integrating out
the angular part we obtain
\[\label{mainresult}
E_0 \left( {\ell ,\mu ,d} \right) = \alpha \left( d \right)\frac{{L^{d - 1} }}{{2^d \pi^{d/2} \Gamma \left( {d/2} \right)\ell ^d }}\int\limits_0^\infty  {dx\,x^{d - 1} \ln \left[ {1 + \frac{{K_1 \left( z \right)}}{{K_2 \left( z \right)}}} \right]}
\]
All along the real axis the function $z$ does not changes its sign and is a function of $q$ and $p_\bot$. This approach to the evaluation of the Casimir energy can be also applied when the integrand is a more complex function of $\Omega_n$. It is convienient to make the replacement $z\to\omega=\sqrt{x^2+\mu^2}$, then we can write equation (\ref{mainresult}) in the alternative way
\[
E_0 \left( {\ell ,\mu ,d} \right) = \frac{{\alpha (d)L^{d - 1} }}{{2^d \pi ^{d/2} \Gamma \left( {\frac{d}{2}} \right)\ell ^d }}\int\limits_\mu ^\infty  {d\omega \,\omega \left( {\omega ^2  - \mu ^2 } \right)^{{\textstyle{d \over 2}} - 1} \ln \left[ {1 + \frac{{K_1 \left( \omega  \right)}}{{K_2 \left( \omega  \right)}}} \right]}
\]
The function $K(\omega)$ can be inferred from the boundary conditions as the next examples show.
\section{Quantum fields confined between parallel surfaces}
Equation (\ref{mainresult}) is a general formula that can be applied to all cases of quantum fields confined between two parallel surfaces on which we impose boundary conditions. A non-trivial example is the one of the Casimir energy for a massive fermionic field confined in a $d+1$-dimensional slab-bag under MIT boundary conditions \cite{Johnson75}, \cite{IJMPA1003}. If the surface of the bag is perpendicular to the $d$-direction MIT boundary cconditions lead to the function
\[
F\left( {p_d \ell } \right) = \mu \sin \left( {p_d \ell } \right) + p_d \ell \cos \left( {p_d \ell } \right)
\]
Hence we can choose $G(\omega)$ as
\[
G\left( \omega  \right) = \mu \frac{{\sin \left( \omega  \right)}}{\omega } + \cos \left( \omega  \right)
\]
Now we write
\[
K\left( \omega  \right) = G\left( {i\omega } \right) = \mu \frac{{\sinh \left( \omega  \right)}}{\omega } + \cosh \left( \omega  \right)
\]
It follows that
\[
K_1 \left( \omega  \right) = \frac{1}{2}\left( {1 - \frac{\mu }{\omega }} \right)e^{ - \omega }
\]

and
\[
K_2 \left( \omega  \right) = \frac{1}{2}\left( {1 + \frac{\mu }{\omega }} \right)e^\omega
\]
The regularised Casimir energy of the fermion field is
\[
E_0 \left( {\ell ,\mu ,d} \right) =  - \frac{{C(d)L^{d - 1} }}{{2^{d - 1} \pi ^{d/2} \Gamma \left( {\frac{d}{2}} \right)\ell ^d }}\int\limits_\mu ^\infty  {d\omega \,\omega \left( {\omega ^2  - \mu ^2 } \right)^{{\textstyle{d \over 2}} - 1} \ln \left[ {1 + \frac{{\omega  - \mu }}{{\omega  + \mu }}e^{ - 2\omega } } \right]}
\]
Setting $d=3$ we obtain the the result given in \cite{MamaevTrunov80}. We can explore this result and lengthy calculation shows that upon expanding the log function we end up with a convergent sum of integrals that can be expressed each one of them as a derivative of the Whittaker function $W_{\nu\mu}$ with respect to an auxiliary variable $\lambda$ at $\lambda=1$. This is as far as we can go analytically. But it is possible to consider and obtain analytical results in the massless limit and the very massive limit, see \cite{IJMPA1003} for details.
\section{Confined bosonic and fermionic fields in a uniform magnetic field}
The analytical regularisation scheme can be also applied to massive bosonic and fermionic fields in a uniform magnetic field perpendicular to the confining surfaces. Let us consider first the case of a charged scalar field under Dirichlet boundary conditions. In this case we will have
\[
E_0 \left( {\ell ,\mu ,eB} \right) = \alpha \left( {\frac{{eB}}{{2\pi }}} \right)\frac{{L^2 }}{2}\sum\limits_{n = 0}^\infty  {\int\limits_0^\infty  {\frac{{dp_3 }}{{2\pi }}\ln \left( {1 - e^{ - 2z} } \right)} }
\]
where we have taken into account that the momenta associated with with the unconstrained directions are related to the Landau levels
\[
p_1^2  + p_2^2  = eB\left( {2n + 1} \right) \;\;\;\; n=0,1,2,3,\dots
\]
The factor $eB/2\pi$ takes into account the degeneracy of the Landau levels. The function $z$ reads
\[
z = z\left( {p_3 ,n} \right): = \sqrt {\ell ^2 p_3^2  + eB\ell ^2 \left( {2n + 1} \right) + \mu ^2 }
\]
where $\mu:=m\ell$. The integrand can be expanded and the Casimir energy can be expressed as a convergent infinite series of integrals  that can be evaluated in terms of the Whittaker functions. The final result reads
\begin{eqnarray}\label{EbosonicB}
E_0 \left( {\ell ,\mu ,eB} \right) &=&  - \frac{{eBL^2 }}{{2\pi ^2 \ell }}\sqrt {eB\ell ^2  + \mu ^2 } \sum\limits_{k = 1}^\infty  {\frac{1}{k}K_1 \left( {2k\sqrt {eB\ell ^2  + \mu ^2 } } \right)} \nonumber \\
&-& \frac{{eBL^2 }}{{2\pi ^2 \ell }}\sum_{n=1}^\infty\,\sqrt {(2n+1)eB\ell ^2  + \mu ^2 } \sum\limits_{k = 1}^\infty  {\frac{1}{k}K_1 \left( {2k\sqrt {(2n+1)eB\ell ^2  + \mu ^2 } } \right)}
\end{eqnarray}
Notice that this is an exact result. If $eB \gg \ell^2$ the behavior
of the Bessel function of the third kind leads us to write
\[
\frac{{E_0 \left( {\ell ,eB} \right)}}{{L^2 }} \approx  - \frac{{\left( {eBL^2 } \right)^{5/4} }}{{\pi ^{1/2} \ell ^3 }}e^{ - 2\sqrt {eBl^2 } }
\]
in agreement with \cite{JPA1999}. For arbitrary $\mu$ and $eB\ell^2$ equation (\ref{EbosonicB}) can be solved numerically, see \cite{JPA2002} for details.

For the case of a massive fermionic field we replace the Dirchlet boundary conditions by MIT bag model ones \cite{Johnson75}. As in the case of the bosonic charged scalar field, analytical regularisation works fine even if the final steps are more much complex than those of the previous case. We state the main results. Starting from equation (\ref{mainresult}) which now reads
\[
E_0 \left( {\ell ,\mu ,eB} \right) =  - 2 \times {\textstyle{1 \over 2}}\left( {{\textstyle{{eBL^2 } \over {2\pi }}}} \right)\sum\limits_{n = 0}^\infty  {\sum\limits_{\alpha  \in \left\{ { - 1,1} \right\}} {I_{n\alpha } } }
\]
where
\[
I_{n\alpha } : = \int\limits_{ - \infty }^{ + \infty } {\frac{{dp_3 }}{{2\pi }}\ln \left[ {1 - \frac{{z - \mu }}{{z + \mu }}e^{ - 2z} } \right]}
\]
with
\[
z = z\left( {q,n,\alpha } \right) = \sqrt {\ell ^2 p_3^2  + \left( {2n + 1 - \alpha } \right)eB\ell ^2  + \mu ^2 } \;\;\;\; n=0,1,2,3,\dots
\]
after expanding the log and summing over $\alpha$ we arrive at
\[\label{EfremionicB}
E_0 \left( {\ell ,\mu ,eB} \right) =  - 2\,\,\frac{{eBL^2 }}{{2\pi ^2 \ell }}\,\,\sum\limits_{p =  - 1}^{\;\;\infty\;\;\;\prime}\,{\sum\limits_{k = 1}^\infty  {\frac{{\left( { - 1} \right)^{k + 1} }}{k}} } \,I_{pk} \left( {M_p } \right)
\]
where the prime means that the term corresponding to $p=-1$ must be multiplied by $1/2$ and
\[
I_{pk} \left( {M_p } \right): = \int\limits_0^\infty  {dx\,\left[ {\left( {x^2  + M_p^2 } \right)^{1/2}  + \mu } \right]^{ - k} \,\,\left[ {\left( {x^2  + M_p^2 } \right)^{1/2}  - \mu } \right]^k } \,e^{ - 2k\left( {x^2  + M_p^2 } \right)^{1/2} }
\]
with $x=p_3\ell$ and $ M_p^2 : = 2\left( {p + 1} \right)eB\ell ^2  + \mu ^2 $ for $p=-1,0,1,2,3,\cdots$.  These integrals are non-trivial ones but a numerical evaluation of equation (\ref{EfremionicB}) is feasible \cite{JPA2002}. Two limiting cases, however can be solved analytically. The first one is limit $\mu\to 0$ and the other one is the limit  $\mu \gg 1$. For the massless limit we obtain
\[
E_0\left(\ell, \mu\to 0,eB \right)\approx -\frac{eBL^2}{48\ell}\;\;\;\; eB\ell^2 \gg 1
\]
and for very massive one
\[
E_0 \left( {\ell ,\mu  \gg 1,eB} \right) \approx  - \frac{{eBL^2 }}{{32\pi ^{3/2} \ell }}\,\,\frac{{e^{ - 2\mu } }}{{\mu ^{1/2} }}
\]
For arbitrary values of $\mu$ and $eB$ numerical analysis leads to very precise results, see \cite{JPA2002} for more details and numerical plots.
\section{ Vacuum energy shift in $\kappa$-deformed theory}
A quantum field theory is $\kappa$-deformed when its spacetime symmetries are described by the $\kappa$-deformed Poincar\'e algebra. These theories which may be important in some models of the early universe lead in general to highly nontrivial dispersion relations. See \cite{CP} and references therein. Consider for example scalar $\kappa$-deformed electrodynamics. The frequency spectrum is given by
\[
\omega \left( p \right) = \sinh ^{ - 1} \left( {\frac{1}{{2\kappa }}\sqrt {p^2  + m^2 } } \right)
\]
The parameter $\kappa$ is a measure of the departure from the Poincar\'e algebra. For Dirichlet boundary coditions imposed on a certain direction, part of the frequency sepectrum is discretised and the vacuum energy shift is given by
\[
E_0 \left( \ell,\eta, m \right) = L^2 \int {\sum\limits_n {\frac{{d^2 p_ \bot  }}{{\left( {2\pi } \right)^2 }}\frac{1}{\eta }\sinh ^{ - 1} \left( {\eta \Omega _n } \right)} }
\]
where as $\Omega_n={\sqrt {p^2_\bot +\lambda^2_n/\ell^2  + m^2 } } $ and we have introduced the parameter $\eta=1/\kappa$.
Making use of the identity
\[
\sum\limits_n {\sinh ^{ - 1} \left( {\eta \Omega _n } \right) =  - \oint {\frac{{dq}}{{2\pi }}} } \sum\limits_n {\frac{{2q}}{{q^2  + \Omega _n ^2 }}\sinh ^{ - 1} \left( {\eta q} \right)}
\]
and integrating by parts
\[
E_0 \left(  \ell,\eta, m  \right) = \frac{{L^2 }}{2}\int {\sum\limits_n {\frac{{d^2 p_ \bot  }}{{\left( {2\pi } \right)^3 }}\oint\limits_{} {\frac{{\mbox{log} \left[ {K\left( z \right)} \right]}}{{\sqrt {1 - \eta ^2 q^2 } }}} } }
\]
Making the decomposition $K\left( z \right) = K_1 \left( z \right) + K_2 \left( z \right)$ and regularising
\[
E_0 \left(  \ell,\eta, m  \right) = L^2 \int {\frac{{d^2 p_ \bot  }}{{\left( {2\pi } \right)^3 }}\int\limits_0^{1/\eta } {\frac{{dq}}{{\sqrt {1 - \eta ^2 q^2 } }}\mbox{log} \left[ {1 + \frac{{K_1 \left( z \right)}}{{K_2 \left( z \right)}}} \right]} }
\]
After integrating the angular part we finally obtain
\[
E_0 \left(  \ell,\eta, m  \right) = \frac{{L^2 }}{{\left( {2\pi } \right)^2 }}\int\limits_0^{1/\eta } {\frac{{dq}}{{\sqrt {1 - \eta ^2 q^2 } }}\,I\left( {q^2 } \right)}
\]
where
\begin{eqnarray}
4\ell^2 \sqrt {q^2  + m^2 }\,I \left( {q^2 } \right) &=& -2\mbox{Li}_2\left(\ex^{-2\ell\sqrt{q^2 + m^2}}\right)\ell q^2-\mbox{Li}_2\left(\ex^{-2\ell\sqrt{q^2 + m^2}}\right)\ell m^2-\mbox{Li}_2\left(\ex^{-2\ell\sqrt{q^2 + m^2}}\right)\ell q^2 \nonumber \\
&-&\sqrt {q^2  + m^2 }\mbox{Li}_2\left(\ex^{-2\ell\sqrt{q^2 + m^2}}\right)
\end{eqnarray}
For $\eta\to 0$ and $m\to 0$ we obtain the standard Casimir energy of a massless scalar field under Dirichlet boundary conditions, For $\eta\neq 0$ and $m=0$ we obtain
\[
E_0 \left(  \ell,\eta, m=0  \right)=-\frac{L^2}{4\pi^2\ell^3}\sum_{n=1}^\infty\frac{1}{n^2}\int_0^{1/\eta} \, dy\left( 1+\frac{1}{2n}\right)\frac{\ex^{-2ny}}{\sqrt{1+\frac{}n^2y^2{\ell^2}}}
\]
in agreement with \cite{CP}.
\section{Photon creation rate for $\kappa$-deformed theory}
In a $\kappa$-deformed theory there is the possibility of quanta creation. The rate  at which photons are created is related to the quantity
\[
S = i\frac{\eta }{{2\pi }}\sum\limits_n {\omega _n^2 }
\]
and the vacuum decay will be proportional to $\exp(-S)$. The analytical regularisation technique can be applied to the evaluation of the sum of the squares of the frequencies which for massless quanta reads
\[
S = i\frac{{\eta L}}{{2\pi }}\int {\frac{{d^2 p_ \bot  }}{{\left( {2\pi } \right)^2 }}\sum\limits_n {\left[ {\frac{1}{\eta }\sinh^{-1} \left( {\eta \sqrt {p_ \bot ^2  + p_n^2 } } \right)} \right]^2} }
\]
We can write
\[
\sum\limits_{n} {\sinh ^{ - 1} \left( {\eta \sqrt {p_ \bot ^2  + p_n^2 } } \right) =  - \sum\limits_{n} {\oint {\frac{{dq}}{{i\pi }}\frac{{q\left[ {\sin ^{ - 1} \left( {\eta q} \right)} \right]^2 }}{{q^2  + \Omega _n^2 }}} } }
\]
then the photon creation rate can be rewritten as
\[
S =  - \frac{{L^2 }}{{16\pi ^4 \eta }}\int {d^2 p_ \bot  \oint {dq} \sum\limits_n {\frac{{2q^2 }}{{q^2  + \Omega _n^2 }}\frac{{\left[ {\sin ^{ - 1} \left( {\eta q} \right)} \right]^2 }}{q}} }
\]
A lengthy and tricky calculation the details of which will be
available elsewhere leads to
\[
S = \frac{L^2}{4\pi^3}\int d^2 p_ \bot \int_{1/\eta}^\infty dq \frac{dq}{\sqrt {1-\eta^2 q^2 }}\mbox{log}\left[1+\frac{K_1(z)}{K_2(z)}\right]
\]
For Dirichlet boundary conditions this is
\[
S = i\frac{{L^2 }}{{4\pi^3 }}\int {d^2 p_\bot}\int\limits_{1/\eta }^\infty  {\frac{dq}{\sqrt{\eta ^2 q^2  - 1}}\mbox{log}\left(1-\ex^{-2\ell\sqrt{p_\bot^2+q^2}}\right)}
\]
We can expand the log and perform the integrals over $p_\bot$  and after some additional manipulations we end up with
\[\label{Sapp}
S = i\frac{{L^2 }}{{4\pi ^2 \ell ^3 }}\sum\limits_{n = 1}^\infty  {\,\frac{1}{{n^2 }}\,\int\limits_{\ell /\eta }^\infty  {\frac{{dy}}{{\sqrt {\left( {\frac{{ny}}{\ell }} \right)^2  - 1} }}\left( {1 + \frac{1}{{2n}}} \right)e^{ - 2ny} } }
\]
If we take the limit $\eta\to 0$ (no deformation) then $S\to 0$. The same happens if we take the limit $\ell\to\infty$. In both limiting cases there is no photon creation. Equation (\ref{Sapp}) is in agreement with \cite{Mendes}.
\section{Conclusions}
In this brief review we derived a general regularised expression for
the evaluation of the Casimir energy of a quantum field in a flat
manifold under the influence of boundary conditions, imposed on the
field on flat surfaces, or topological conditions constraining the
motion along a particular spatial direction. We exemplified the main
result with the case of a massive fermionic quantum field confined
by a planar \( d+1 \) dimensional slab-bag with MIT boundary
conditions and confined bosonic and fermionic fields under a uniform
magnetic field. We also sketched the calculation of the vacuum
energy and photon creation in $\kappa$-deformed theories.

\end{document}